\documentclass[12pt]{iopart}
\usepackage{graphicx}

\begin{document}

\title{Vortex dynamics as a function of field orientation in BaFe$_{1.9}$Ni$_{0.1}$As$_2$ }

\author{ S. Salem-Sugui, Jr.$^1$, L. Ghivelder$^1$, A.D. Alvarenga$^2$, L.F. Cohen$^3$, Huiqian Luo$^4$, Xingye Lu$^4$}

\address{$^1$Instituto de Fisica, Universidade Federal do Rio de Janeiro,
21941-972 Rio de Janeiro, RJ, Brazil}
\address{$^2$Instituto Nacional de Metrologia Qualidade e Tecnologia, 25250-020 Duque de Caxias, RJ, Brazil}
\address{$^3$The Blackett Laboratory, Physics Department, Imperial College London, London SW7 2AZ, United Kingdom}
\address{$^4$Beijing National Laboratory for Condensed Matter Physics, Institute of Physics, Chinese Academy of Sciences, Beijing  100190,  P. R. China}
\begin{abstract}
Vortex dynamics in a multiband anisotropic superconducting such as the Fe based superconductors, is interesting and potentially important for applications. In this study we examine flux-creep data for fields along the second magnetization peak  observed in $M(H)$ curves of BaFe$_{1.9}$Ni$_{0.1}$As$_2$ for  $H$$\parallel$$c$-axis, $H$$\parallel$$ab$-planes and $H$ forming a 45$^o$ angle with $ab$-planes.  We find that  the $M$-$H$ loops taken from the different field directions can be collapsed onto a single universal curve at all temperatures with a simple scaling factor equivalent to the superconducting anisotropy, showing not only that the vortex pinning is isotropic, three dimensional and most likely related to point like defects. The resulting critical currents however, taken from the Bean model appears to show enhanced low field pinning for $H$$\parallel$$c$. The features in the vortex- dynamics also differ in different field orientations and show no direct correlation with the second magnetization peak $Hp$ as is the case with a direct crossover in pinning regimes. Isofield plots of the scaled activation energy obtained from flux-creep data is found to be a smooth function of temperature as the $Hp(T)$ line is crossed consistent with a single type of pinning regime operating at this field, independent of field orientation. The functional form of the $Hp(T)$ lines in the resulting phase diagrams also support this view.
\end{abstract}
\maketitle

\section{Introduction}
The study of vortex-dynamics in the novel pnictide superconductors\cite{1} has attracted increasing attention \cite{2,3,4,5,6,7,8,9,9b,9c}, due to the considerable high-$T_c$ of these compounds, when compared to the conventional superconductors, and because of similarities with the high-$T_c$ cuprates superconductors \cite{norman}. As in the cuprates\cite{yeshurun,gautam}, most of the pnictides systems exhibit the second magnetization peak, or fish-tail, in isothermic magnetization curves, as well as large flux-creep, allowing the study of different regions of the vortex-phase diagram in certain detail. The second magnetization peak is associated with a maximum in the critical current when measured as a function of field at a fixed temperature. This phenomena is not yet completely understood in pnictides\cite{4}. Although pnictides have similarities with the cuprates, such as the layered structure and antiferromagnetism of the precursor non-superconducting system \cite{norman}, it is well established that superconductivity in pnictides has a multi-band character\cite{tesa,ding} for which it is predicted the existence of non-Abrisosov vortices\cite{10,11,12} potentially leading to new effects in the vortex matter. In this work we study the vortex dynamics in a BaFe$_{1.9}$Ni$_{0.1}$As$_2$ superconductor single crystal, by means of isofield magnetization $M(H)$ curves and magnetic relaxation $M(t)$ curves obtained for  $H$$\parallel$$c$-axis, $H$$\parallel$$ab$-planes and $H$ forming a 45$^o$ angle with $ab$-planes. .
\section{Experimental} 
Magnetization data were obtained by using commercial magnetometers: a 5T MPMS based on a superconducting quantum interference device (SQUID) was used for most of measurements with $H$$\parallel$$ab$-planes ; and a 9T PPMS was used for the other measurements including all data with $H$$\parallel$$c$-axis and and $H$-45$^o$-$ab$-planes. The measurements were made after lowering the sample temperature from above $T_c$ in zero applied magnetic field (ZFC-procedure). The studied sample, is a high-quality single crystal of BaFe$_{1.9}$Ni$_{0.1}$As$_2$ with transition temperature $T_c$ = 20 K, transition width $\Delta$$T_c$=0.3 K, mass = 120 mg and dimensions 0.5 x 1.6 x 0.02 cm. Details of the sample preparation can be found in Ref. \cite{luo}. For $H$$\parallel$$c$-axis and $H$-45$^o$-$ab$ geometries, we carefully broke the sample and used a 43.1 mg piece with dimensions ~0.55x0.5x0.02 cm. For  $H$$\parallel$$ab$-planes, the sample was attached to a hard-plastic slab perfectly inserted along the entire length of the straw tube used in the measurement systems, ensuring field alignment to within 2 degrees with the $ab$-planes. 

\section{Results and discussion}
\begin{figure}[t]
\includegraphics[width=\linewidth]{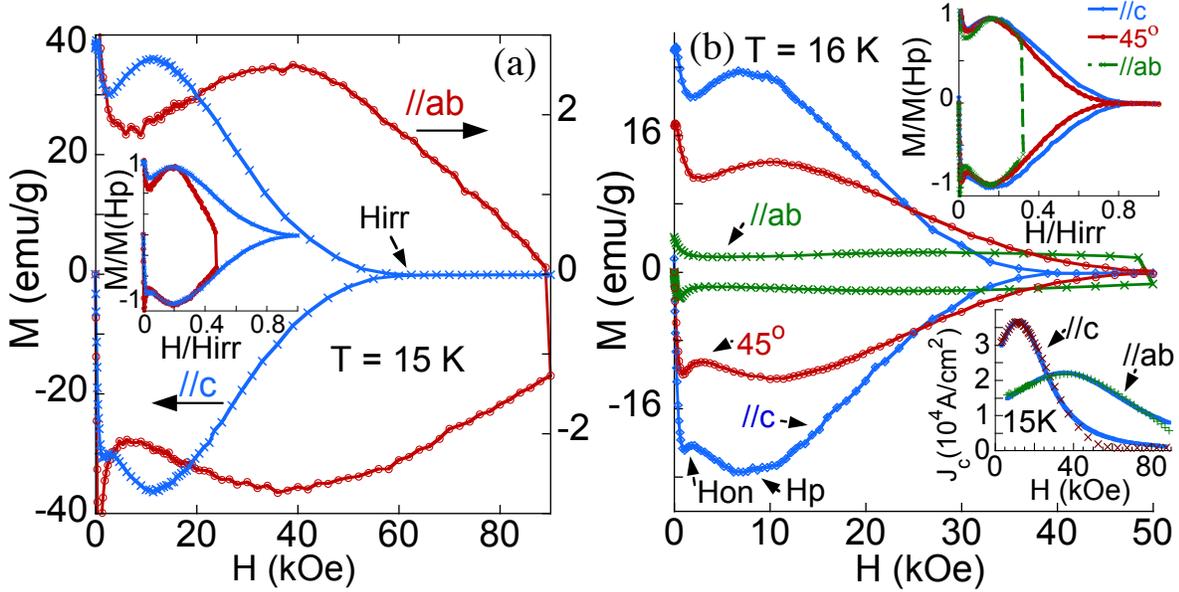}
\caption{a) double plot of $M(H)$ at 15 K for $H$$\parallel$$c$ and $H$$\parallel$$ab$.; b) $M(H)$ curves at 16 K for the three geometries. Insets: a) and upper b) shows a reduced plot of the same correspondent $M(H)$ curves.; lower b), $J_c(H)$ at 15 K for $H$$\parallel$$c$ and $H$$\parallel$$ab$. Solid lines represent a fitting to the data.}
 \label{fig1}
\end{figure}
Figure 1a shows a double plot of $M(H)$ curves obtained at 15 K for $H$$\parallel$$c$ and $H$$\parallel$$ab$ where it is possible to observe the differences in the shape of the fish-tail between both field directions, which as shown below, are due to differences in the vortex dynamics. Figure 1b shows $M(H)$ curves at 16 K for all field directions, all exhibiting the second magnetization peak, where $Hon$ represents the onset of the second magnetization peak with its maximum value occurring at $Hp$. The inset of Fig. 1a and the upper inset of Fig. 1b show a reduced plot of the curves appearing in the respective main figures, where $M$ is normalised by $M(Hp)$ the value of magnetization at the respective $Hp$ and equivalently the $H$ scale by the respective irreversible field $Hirr$. These inset figures show that the $M$-$H$ loops from the different field orientations can be collapsed onto one universal curve and therefore the same underlying physics is occurring independent of the orientation of field.  If we compare $Hirr$($H$$\parallel$$ab$)/$Hirr$($H$$\parallel$$c$), $Hp$($H$$\parallel$$ab$)/$Hp$($H$$\parallel$$c$) we find a factor that is of a value similar to the reported underlying anisotropy of the coherence length \cite{abc}. So the pinning properties are governed by the superconducting anisotropy. Although this appears straightforward, when we compare the absolute value of the critical current density from the different field directions (by using the Bean Model \cite{bean} and the crystal dimensions above listed), as shown in the lower inset of Fig. 1b for T = 15 K, we find that at low field, $J_c$($H$$\parallel$$c$) exceeds  $J_c$($H$$\parallel$$ab$). Interestingly, this unusual effect in the critical current, is predicted in Ref. \cite{9c} for ÓmodestÓ anisotropic systems with single pinning centers (acting as point-like pinning) which are physically large compared to the coherence length exhibiting strong pinning for fields applied parallel to the c-axis. This appears to be the case for low fields, implying that althought the field scales with the underlying anisotropy, the complexity of the multiband nature of the superconductor is evident in the absolute magnitude of the critical current density. In more anisotropic systems it is relevant to consider the 3D simulation of vortices interacting with single pinning centers for a fixed magnetic field presented in Refs. \cite{olson1,olson2}, showing that an anisotropy dependent transition from 3D to a quasi 2D line vortex appears associated with an increase in the critical current. Although in the system we consider here with much more modest anisotropy a 3D to 2D crossover is not directly relevant.
The lower inset of Fig. 1b  shows fittings of $J_c(H)$ at 15 K to the expression $J_c(B)$$=$$A/[(B-Bp)^2+(\Delta B)^2]^{5/4}$ of Ref. \cite{rosenstein} where $A$, $Bp$ and $\Delta B$ are fitting parameters, where $Bp$ is the peak position and $\Delta B$ is the peak width. The resulting fittings conducted in a wide field range are excellent suggesting that the shape of the $J_c(H)$ curve is primarily governed by vortex lattice softening as a function of field. The fitting shown in the lower inset of Fig. 1b produced $Bp$ = 11.7 kOe and $\Delta B$ = 20 kOe for $H$$\parallel$$c$ and $Bp$ = 35.3 kOe and $\Delta B$ = 47 kOe. Similar fittings were obtained at different temperatures.

The almost perfect symmetry of the $M(H$) curves of Fig. 1 with respect to the $x$-axis  is evidence that bulk pinning is dominant. Also, the equilibrium magnetization, $Meq$, defined as the average value of $M$ on both branches of a $M(H)$ curve, is very small, so we may use values of $M$ instead ($M-Meq$) in the analysis that follows. Near $T_c$ for temperatures above 19 K, the second magnetization peak is no longer observed. 

In order to understand whether a vortex lattice softening model is sufficient to explain all features of the data, a more detailed vortex dynamics study was performed by collecting isofield magnetic relaxation data, $M(t)$ curves, along several isothermic $M(H)$ curves, for fields lying below and above the second magnetization peak, for the three geometries. We also obtained isofield magnetic relaxation curves as a function of temperature. Magnetic relaxation data were collect for 2 hrs when obtained in both branches of $M(H)$ curves and for 3.5 hrs when only in the lower branch. We also measured long time magnetic relaxation for approximately 12 hrs in different regions of the $M(H)$ curve for $H$$\parallel$$c$-axis. All $M(t)$ vs. log$(t)$ curves show strictly linear behavior, starting above a transient time $\tau_0$ $\approx$ 3-4 min. Such a large transient time was observed before in BaFe$_{1.82}$Ni$_{0.18}$As$_2$ with $T_c$=8 K, and seems to be intrinsic of the material \cite{8}. The linear behavior with log$(t)$  was also observed for the 12 hrs relaxation curves. This fact suggests that it is more appropriate to analyse the data by using the relaxation rate $R$ = $dM/dln(t)$ as defined in Ref. \cite{beasley}. As shown in Ref. \cite{beasley} one may obtain information on the apparent activation energy, but this quantity may not have any physical meaning in our data, since the magnetization $M_0$ at the time $t$ = 0, above which logarithmic relaxation should start, is not well defined, due to the $\approx$ 4 min long transient region (see Fig. 2a below).
    \begin{figure}[t]
\includegraphics[width=\linewidth]{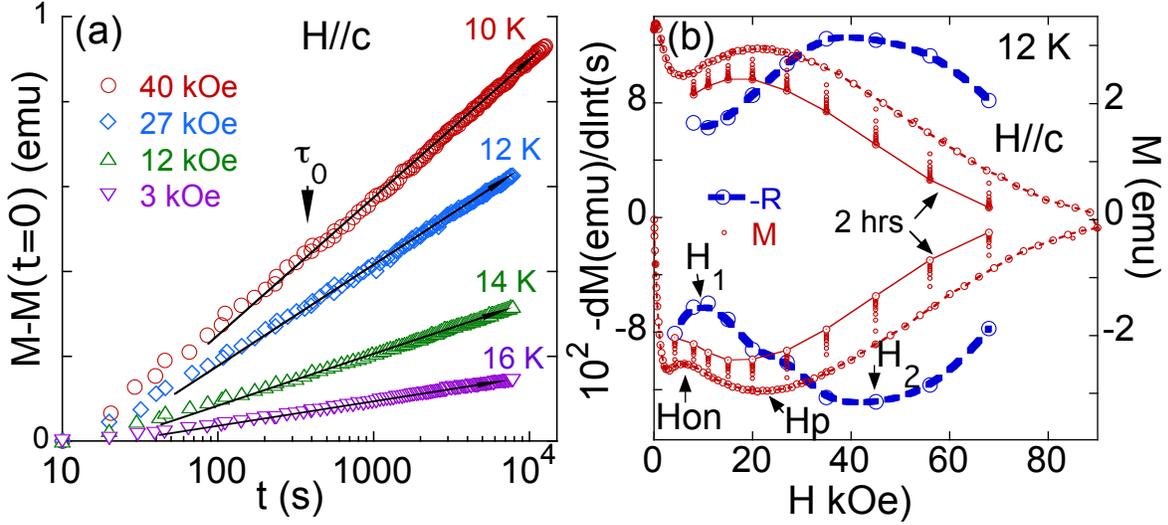}
\caption{$H$$\parallel$$c$-axis: a) selected magnetic relaxation data  b)Double plots of $-R~vs.~H$ and $M(t)~vs.~H$ at  $T$ = 12 K} 
 \label{fig2}
\end{figure}
\begin{figure}[t]
\includegraphics[width=\linewidth]{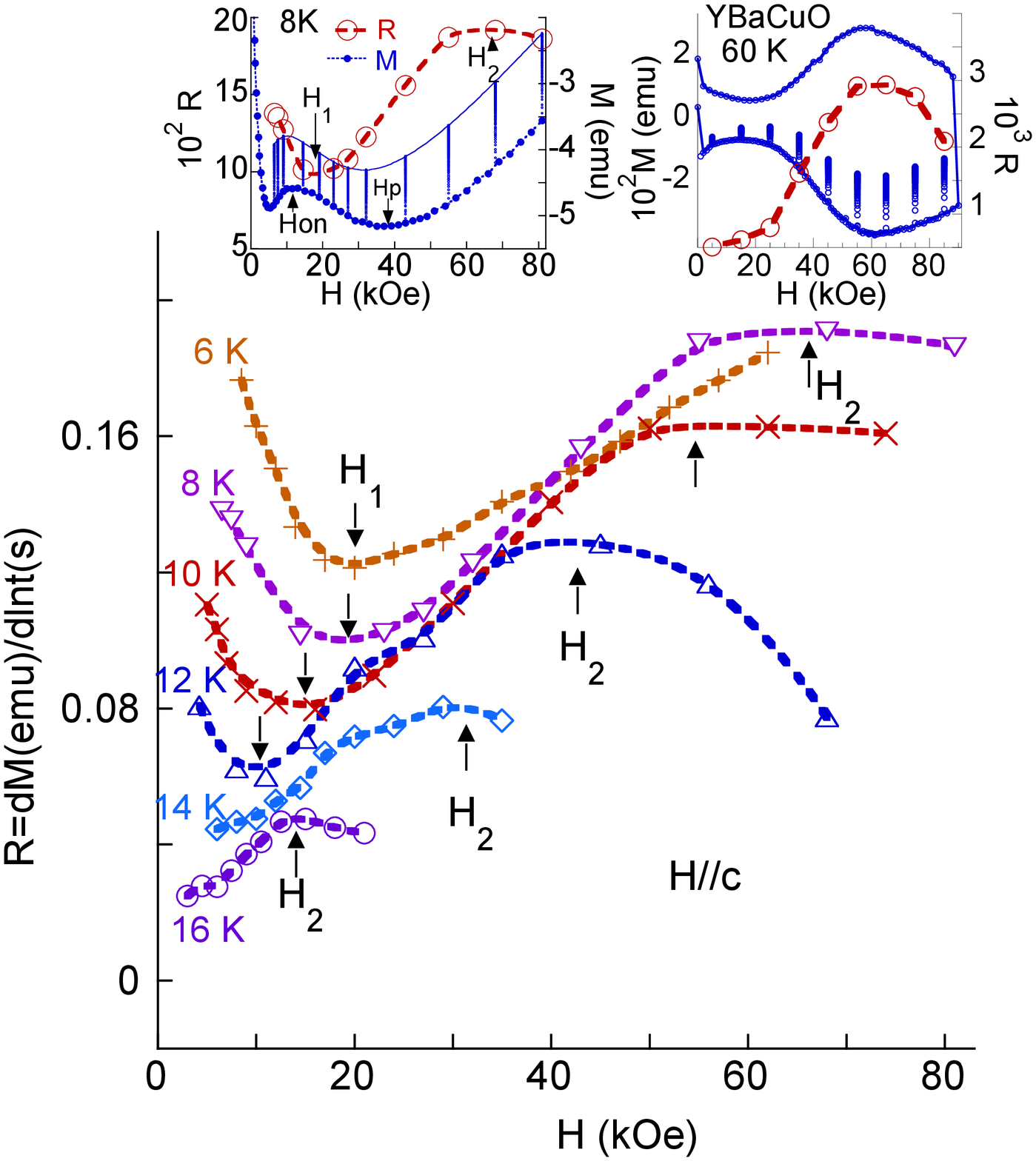}
\caption{$H$$\parallel$$c$-axis: $R~vs.~H$ for the lower branch of $M(t,H)$ curves. Upper insets: left) Double plots of $R~vs.~H$ and $M(t)~vs.~H$ at $T$ = 8 K.  right) Double plot of $R~vs.~H$ and $M(t,H)$ for YBaCuO at $T$ = 60 K.}
 \label{fig3}
\end{figure}

Figure 2 shows results of the analysis of $M(t)$ data for $H$$\parallel$$c$-axis. The logarithmic relaxation of the magnetization, is exemplified by the selected curves plotted in Fig. 2a.
Figure 2b shows a double plot of $-R(H)~vs.~H$ and the corresponding $M(H)$ curves where the 2hrs relaxation data are also plotted. In Fig. 2b, we plot  -$R$ as a function of increasing and decreasing field and we define $H_1$ as the minimum in $R$ ($R$ increases above $H_1$) and $H_2$ a maximum ($R$ decreases above $H_2$). These two fields, $H_1$ and $H_2$, are defined from data obtained in the incresing field branch of $M(H)$ curves. It is interesting to note that the resulting -$R(H)$ curve resembles the correspondent $M(H)$ curve after a shift to the right. The same trend as shown in Fig. 2b was observed for $T$ = 14 K and 16 K data. 

Figure 3 shows the results of $R~vs.~H$ for relaxation data obtained on the lower branch of all $M(H)$ curves, where the arrows pointing up and down show the approximately positions of $H_2$ and $H_1$ respectively. The position of $Hp$ in each curve of Fig. 3 (not shown) lies approximately at  ($H_1$+$H_2$)/2. It is clear from Figs. 2b and 3, which are representative data from all experiments, that there is no apparent change in the relaxation rate as the peak field $Hp$ is crossed. For the sake of comparison, the upper inset of Fig. 3 shows double plots, as in Fig. 2b, of data obtained at the lower branch of an $M(H)$ curve at 8K for $H$$\parallel$$c$ (left inset) and data obtained in a YBaCuO sample ($T_c$ $\approx$ 92 K) (right inset), with relaxation data obtained during 60 min ($H$$\parallel$$c$-axis) at the lower branch of an $M(H)$ curve at $T$ = 60 K. A perfect matching between $Hp$ and the field position of the maximum in $R$ is clear for YBaCuO, which in this case represents a pinning crossover\cite{abulafia} taking place as $Hp$ is crossed. A direct comparison of these two figures shown in the upper inset of Fig. 3, suggests that the second magnetization peak in these two systems arise from different mechanisms.  
\begin{figure}[t]
\includegraphics[width=\linewidth]{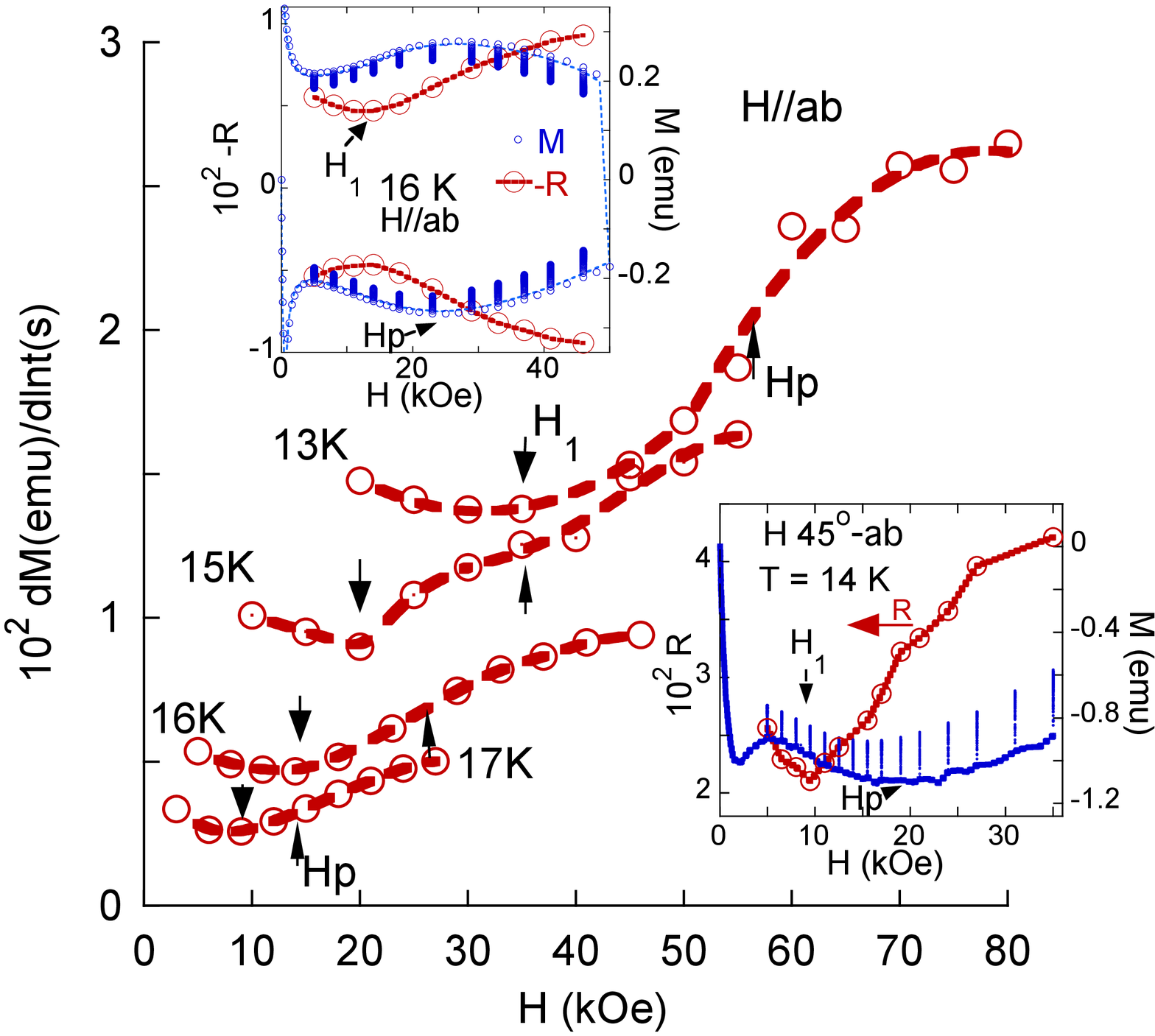}
\caption{$R~vs.~H$ for the lower branch of $M(t,H)$ curves for $H$$\parallel$$ab$-planes. Insets: upper) Double plots of $-R~vs.~H$ and $M(t,H)$ at $T$ = 16 K with $H$$\parallel$$ab$; lower) Double plots of $R~vs.~H$ and $M(t,H)$ at $T$ = 14 K with $H$45$^o$$ab$.}
 \label{fig4}
\end{figure}

Figure 4 shows  the results for $H$$\parallel$$ab$-planes. In that case $R$ also decreases as field increases above $Hon$, reaching a minimum at $H_1$, which lies well below $Hp$. For each curve of Fig. 4, arrows pointing up shows the position of $Hp$ and pointing down of $H_1$. A similar result was previously observed for an overdoped BaFe$_{1.82}$Ni$_{0.18}$As$_2$ with $T_c$ = 8 K for $H$$\parallel$$c$-axis and $H$$\parallel$$ab$-planes\cite{8}. The upper inset of Fig. 4 shows a double plot of -$R~vs.~H$ and the correspondent $M(t,H)~vs.~H$ curve, exhbiting a kind of a shift to the right effect as observed in Fig. 2b for $H$$\parallel$$c$. It is interesting to note that the field $H_1$ for $H$$\parallel$$ab$ is more separated from the field $Hon$ when compared to the $H$$\parallel$$c$-axis case. To check whether this change in $R$ only occur when $H$$\parallel$$ab$-planes, we measure the sample for an intermediate geometry, with $H$ forming a 45$^o$ with $ab$-planes. The results for $T$ = 14 K is shown in the lower inset of Fig. 4 (the same trend is observed for $T$ = 16 K), showing that the position of the field $H_1$ relatively to $Hon$ for this geometry is similar to that observed for $H$$\parallel$$ab$-planes. One may associate the differences in the relaxation rate (as in the position of the field $H_1$ relatively to $Hon$) as the field rotates away from the $c$-axis (where normalised relaxation rate $S$ = (1/$M$)$R$ shows that the creep rate is approximately three times faster for $H$$\parallel$$ab$, than $H$$\parallel$$c$), with the unexpected behavior of the critical current observed with the same rotation, as shown in the lower inset of Fig. 1b, where  $J_c$($H$$\parallel$$c$) exceeds  $J_c$($H$$\parallel$$ab$ at low fields).
\begin{figure}[t]
\includegraphics[width=\linewidth]{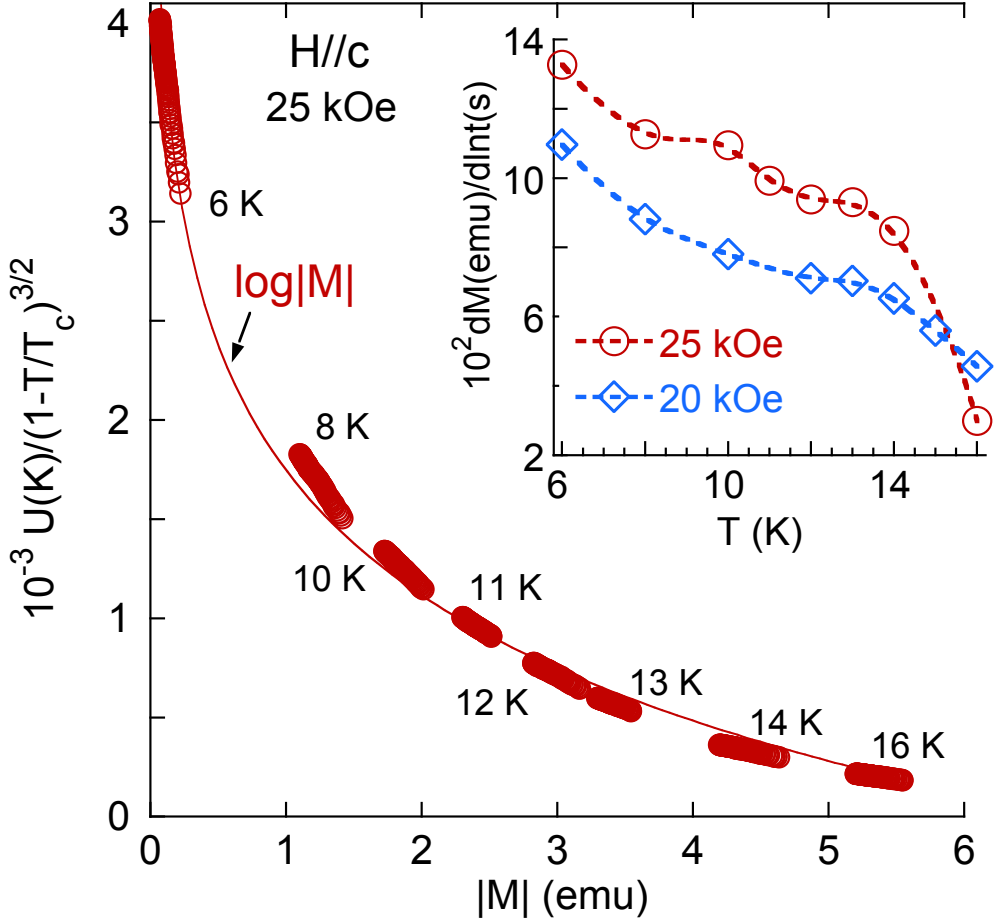}
\caption{$U(M,T)/(1-T/Tc)^{3/2}~vs.~ |M|$ for $H$ = 25 kOe, $H$$\parallel$$c$-axis. The solid line represents a $log|M|$ curve. Inset: $R~vs.~T$ for $H$ = 20 and 25 kOe, $H$$\parallel$$c$-axis.}
 \label{fig5}
\end{figure}

To check whether a pinning crossover, or a vortex phase transition, would perhaps become evident near $Hp$ using a different approach, we obtained a systematic set of isofield magnetic relaxation data as a function of temperature for  $H$$\parallel$$c$ and  $H$$\parallel$$ab$. Now, as in Ref.\cite{maley2}, we calculate the pinning activation energy for a set of isofield data $M(t,T)$ using the expression, $U(M)$=-$T$ln($dM/dt$)+$CT$ where $C$=ln($Bwa/\pi d$) is a constant, where $B\approx H$ is the magnetic induction, $\omega$ is an attempt frequency, $a$ is the hop distance and $d$ is the sample size.  
It is believed\cite{maley2} that the isofield $U(M,T)/g(T/T_c)$ (where $g(T/T_c)$ is an appropriated scaling function of $U$\cite{maley2}) should be a smooth function of $|M|$ within a temperature region with same pinning mechanism.  Figure 5 shows the results of $U(M,T)/g(T/T_c)$ plotted against $|M|$ for $H$$\parallel$$c$, as obtained for $H$ = 25 kOe data with $C$ = 10 and $g(T/T_c)$ = $(1-T/T_c)^{3/2}$. Similar values of the constant $C$ were found for cuprates superconductors\cite{maley2} and for pnictides\cite{4,8}. The almost perfect log$|M|$ fit linking all the data in Fig. 5 suggests the existence of only one pinning mechanism over the entire $\Delta T$ range, for which $Hp$ is located near 11 K.  The same trend was observed for $U(M,T)/(1-T/T_c)^{3/2}~vs.~ |M|$ with $H$ = 20 kOe and $C$ = 10 for $H$$\parallel$$c$, and for $H$$\parallel$$ab$ data with $H$ = 15 kOe where instead $g(T/T_c)$, we used $g(T/T_{irr})$\cite{maley2} with $T_{irr}(15kOe)$=19 K  and $C$ = 13  (not shown). The inset of Fig. 5a shows plots of $R~ vs.~ T$ as obtained from isofield data with $H$ = 20 and 25 kOe for $H$$\parallel$$c$, which do not show any visible effect near $Hp$ located at ~12 K and ~11 K in each respective curve. One would expect some feature in the plot of $R~vs.~T$ as $Hp$ is crossed, either for a pinning crossover or for a vortex-lattice phase transition. The same behavior was observed on similar plots for $H$$\parallel$$ab$-planes with $H$ = 15 and 20 kOe. 

\begin{figure}[t]
\includegraphics[width=\linewidth]{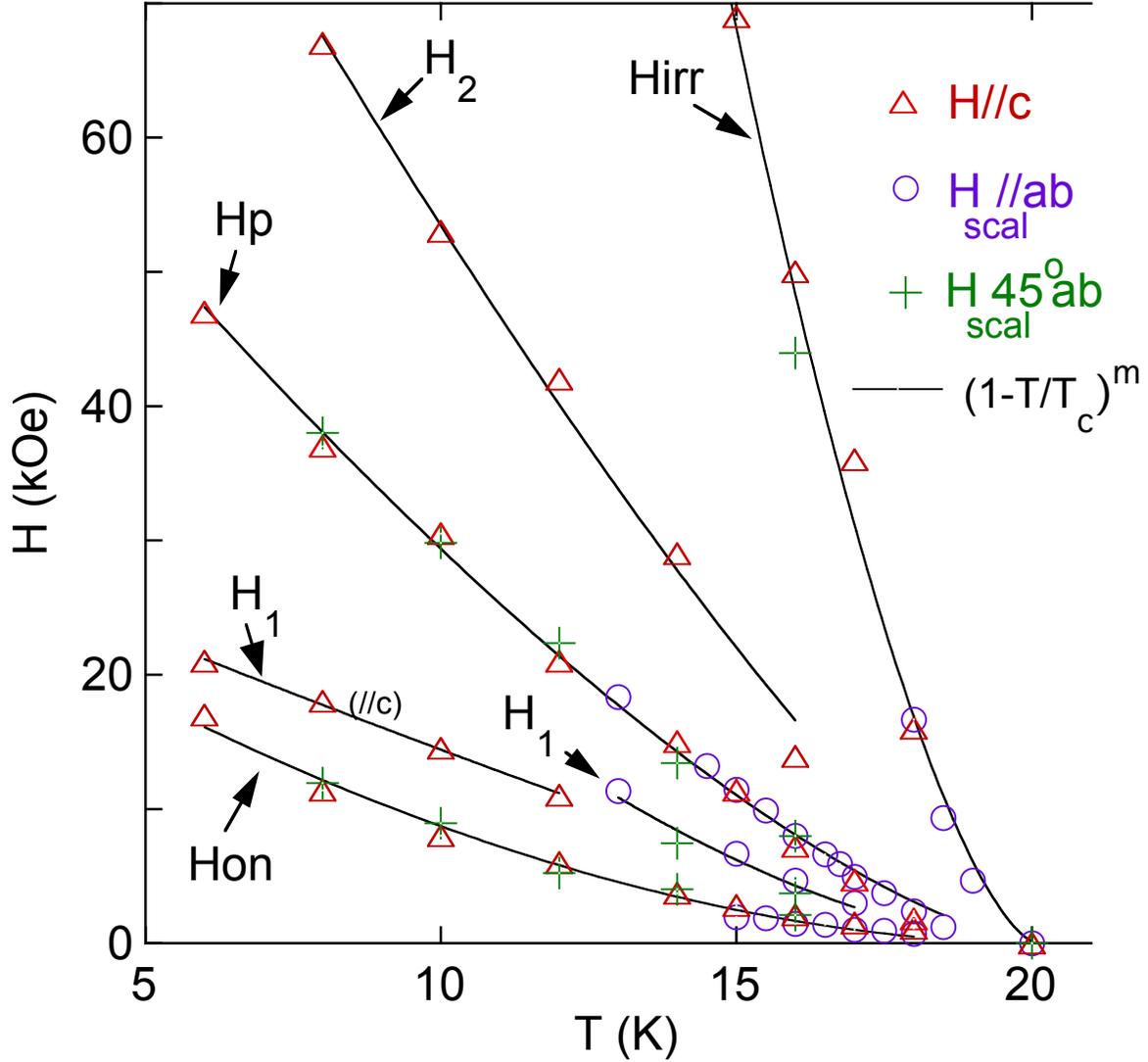}
\caption{$H$vs.$T$ diagram for the three geometries after scale the Y-axis for $H$$\parallel$$ab$ and $H$45$^o$$ab$. Solid lines represents a fitting of the data.}
 \label{fig5}
\end{figure}
Figure 6 shows a $H$vs.$T$ diagram where characteristic fields of the three geometries are plotted against temperature after being divided (scaled) by $\sqrt {(sin(\theta ))^2+(1/3)(cos(\theta ))^2}$ where $\theta$ is the angle between $H$ and the $ab$-plane, and the factor 3$\approx$$Hp$($H$$\parallel$$ab$)/$Hp$($H$$\parallel$$c$)$\approx$$Hirr$($H$$\parallel$$ab$)/$Hirr$($H$$\parallel$$c$) is of the order of the system (upper critical field) anisotropy  \cite{abc}. The collapse of the $Hon$, $Hirr$ and $Hp$ lines is evident and suggests that the same underlying physics is occurring independent of the orientation of field. Each solid line in Fig. 6 follows a $(1-T/T_c)^m$ dependence with $m$ = 1.8 ($Hon$), 1.6 ($Hirr$), 1.4 ($Hp$), 1.7 ($H_1$ for $H$$\parallel$$ab$ and $H$45$^o$$ab$), 1.2 ($H_1$ $H$$\parallel$$c$) and 1.3 ($H_2$). It is interesing to note the differences in the temperature behavior of the $H_1$ lines in Fig. 6. Figure 6 also suggests a collapse of $H_1$ values ( for $H$$\parallel$$ab$ and $H$45$^o$$ab$) which lie between $Hon$ and $Hp$. The feature in $R$ associated with the field $H_1$ appears to be intrinsic of the studied system, since it was also observed in an overdoped sample of the same system \cite{8}. Values of $H_2$ (for $H$$\parallel$$c$) lie between  $Hp$ and $Hirr$. One may associate the anisotropic vortex dynamics observed here with the anisotropic neutron spin resonance found on a similar optimally doped sample\cite{luo2}. Interestingly, isotropic vortex dynamics\cite{8} and isotropic neutron spin resonance\cite{luo3} were found on overdoped samples of BaFe$_{2-x}$Ni$_x$As$_2$. Although evidence suggests no crossover in pinning regime associated with $Hp$, the peak field position $Hp$ is time dependent, because of the difference of relaxation rate as a function of field. The time dependence of $Hp$ is also expected for crossover from elastic to plastic pinning regimes \cite{abulafia}, but as expected the $Hp$ line we measure here does not fit the predicted elastic to plastic crossover supporting our claim that the second magnetization peak in this system is unrelated to this particular mechanism. Nevertheless the fittings of $J_c(H)$ shown in the inset of Fig. 1b suggest that the second peak occurs due to softening of the vortex lattice occurring near $Hp$. This scenario, would accommodate the fact that the rate of relaxation increases with field above $Hp$ (as shown in Figs. 2, 3 and 4 for all geometries) as is expected for example in the case of a disordered vortex phase \cite{5}. In that case, $R$ should eventually decay as the magnetic field approaches $Hirr$ which explains why the field $H_2$ occurs above $Hp$, differently than in YBaCuO (see the upper inset of Fig. 3).  It should be mentioned that it seems that there are few reports in the literature on vortex  phenomenology with respect to  the system studied here, but experiments conduct on BaFe$_{1.8}$Co$_{0.2}$As$_2$ indeed show a disordered vortex phase for intermediate-high fields \cite{yin}.

\section{Conclusions}
In conclusion, our study of vortex dynamics shows several interesting observations. The field scales governing the shape of the $M(H)$ loops scale with the underlying superconducting anisotropy. The shape of the $J_c(H)$ curve exhibiting the peak effect, can be described in terms of vortex lattice softening as a function of increased magnetic field. The differences in the relaxation rate with field orientation and the field anisotropy of $J_c$ indicates that the pinning mechanism may be related to single ("point-like")  pinning centers which are physically large compared to the coherence length \cite{9c} for $H$$\parallel$$c$, acting as strong pinning at low fields. No change in pinning regime is identified as a function of field near the second magnetization peak of $M(H)$ curves in BaFe$_{1.9}$Ni$_{0.1}$As$_2$ either from the functional form of the $Hp(T)$ curve or from the isofield activation energies. 
\section*{Acknowledgements}
SSS, LG and ADA acknowledge support from CNPq and FAPERJ, LFC thanks the UK Funding Council the EPSRC grant EP/H040048.

\end{document}